\documentclass[reprint,superscriptaddress,amsmath,amssymb,aps,prb]{revtex4-1}

\usepackage{graphicx}
\usepackage{graphics}
\usepackage{amsmath}
\usepackage{dcolumn}
\usepackage{bm}
\usepackage{subfig}

\begin{document}


\title{Photon-mediated electronic correlation effects in irradiated two-dimensional Dirac systems}

\author{Jin-Yu Zou}
\affiliation{Institute of Physics, Chinese Academy of Sciences, Beijing 100190, China.}
\affiliation{School of Physical Sciences, University of Chinese Academy of Sciences, Beijing 100190, China}
\author{Bang-Gui Liu}%
 \email{bgliu@iphy.ac.cn}
\affiliation{Institute of Physics, Chinese Academy of Sciences, Beijing 100190, China.}
\affiliation{School of Physical Sciences, University of Chinese Academy of Sciences, Beijing 100190, China}

\date{\today}

\begin{abstract}
Periodically driven systems can host many interesting and intriguing phenomena. The irradiated two-dimensional Dirac systems, driven by circularly polarized light, are the most attractive  thanks to intuitive physical view of the absorption and emission of  photon near Dirac cones. Here, we assume that the light is incident in the two-dimensional plane, and choose to treat the light-driven Dirac systems  by making a unitary transformation to capture the photon-mediated electronic correlation effects, instead of using usual Floquet theory. In this approach, the electron-photon interaction terms can be cancelled out and the resultant effective electron-electron interactions can produce important effects.
These effective interactions will produce a topological band structure in the case of 2D Fermion system with one Dirac cone, and can lift the energy degeneracy of the Dirac cones for graphene. This method can be applicable to similar light-driven Dirac systems to investigate photon-mediated electronic effects in them. 
\end{abstract}

\pacs{Valid PACS appear here}
\maketitle


\section{\label{intro}Introduction}

Periodically driven systems attract more and more attention because of their novel phenomena which are absent in their corresponding equilibrium states\cite{1,2,3,4,5,6,7,8,9,10,11,12,13,14,15,16,17}. The driving can be realized by applying periodical external electric/magnetic fields, by adjusting the parameters of amplitude and frequency of light, and so on.  It can produce various interesting phases such as Chern insulator\cite{5,6,18}, Weyl semimetal\cite{11,16,17}, photovoltaic Hall effect\cite{1,3}, and Quantum Floquet anomalous Hall states and quantized ratchet effect\cite{a1}, and light field can drive large currents in graphene\cite{addn}.

These novel properties are demonstrated by the Floquet operator in terms of the Floquet theorem\cite{19,20,21}, in which the eigen-states of the time-translation-invariant Hamiltonian $\mathcal{H}(x,t)=\mathcal{H}(x,t+T)$ can be written as $|\Psi_{\alpha,\mathbf{k}}(\mathbf{x},t)\rangle=e^{i\mathbf{k}\cdot\mathbf{x}-i\varepsilon_{\alpha,\mathbf{k}}t}|u_{\alpha,\mathbf{k}}(\mathbf{x},t)\rangle$, where $\varepsilon_{\alpha,\mathbf{k}}$ stands for the quasi-energy of the Floquet state, $\alpha$ is the band index and $|u_{\alpha,\mathbf{k}}(\mathbf{x},t)\rangle$ is periodic in $\mathbf{x}$ and $t$. The quasi-energy $\varepsilon_{\alpha,\mathbf{k}}$ is $\omega$ periodic with $\omega=2\pi/T$. The Floquet method can describe the photon absorption and photon emission as well, and has been used to study the exotic phenomena in graphene irradiated by circularly polarized light\cite{1,6,10,15}, but it is not suitable when the light travels in the graphene plane. In this situation, the electrons in graphene feel the light as a linearly polarized one, and are not willing to absorb photon because the Dirac cone of the low-energy bands are spin locked\cite{22}. This can be made easy to understanding as follows. For a Dirac Hamiltonian $\mathcal{H}=\mathbf{k}\cdot\sigma$, where $\sigma$ is Pauli matrix, its eigen states are determined by $\mathcal{H}|\psi_\pm \rangle=\epsilon_\pm |\psi_\pm \rangle$, and it will be easy to derive the spin expectation:
\begin{equation}\label{lock}
\begin{split}
    \langle\psi_\pm|\sigma_j|\psi_\pm \rangle&=\frac{1}{2\epsilon_\pm}\langle\psi_\pm|(\sigma_j\mathcal{H}+\mathcal{H}\sigma_j|)\psi_\pm \rangle\\
                                             &=k_j/\epsilon_\pm
\end{split}
\end{equation}
where $j$ can be $(x,y)$ in the 2D cases or $(x,y,z)$ in the 3D cases. Because the two bands of a Dirac cone have opposite spins, one needs a photon with the angular momentum of either $+1$ or $-1$ to excite an electron from the lower band to the upper band at almost the same k points, but it is impossible in this situation. Therefore, the circular-polarized light travelling in the graphene plane cannot cause direct electron transition near the Dirac cones.

Although it is the case for the direct electron-photon interaction, virtual photon processes of absorption and emission can cause effective second-order electron-electron interactions. Considering these, here, we choose to treat the external field travelling in the 2D plane by using the second quantization form, and thereby study effective electronic interactions, like those from the electron-phonon interaction in superconductors. For a 2D Fermion system with one Dirac cone, the effective interactions will produce a topologically nontrivial band structure with unavoidable band-crossing[27]. Applied to graphene, the key finding is that the Dirac cones remain gapless, but the energy degeneracy of the Dirac cones can be lifted when one adjusts the frequency and amplitude of electromagnetic waves properly. These unexpected interesting results can be attributed to the quantum effects of the irradiated electron systems when the key quantum photon-electron interaction is captured. Moreover, this method can be applicable to similar systems.

\section{Model}

The initial Hamiltonian that describes a 2D Dirac fermion in a graphene-like system can be expressed as
\begin{equation}\label{dirac}
  H_D=\sum_kc_k^\dagger k_i\sigma^i c_k  \quad i=x,y
\end{equation}
where $\vec{k}$ is the momentum in the 2D plane, $\sigma^i$ denotes the Pauli matrix, and $c_k$ is the two-component annihilation operator with pseudo-spin. We use the summation convention: the same index means summation over it. We apply a light on the 2D fermion system in the y direction in the 2D plane, as shown in Figure 1. This in-plane light driving is different from those with perpendicular incident light\cite{addn}. The vector potential of the light lies in the x-z plane, which means that only the x component distributes to the coupling with the Dirac fermions.

\begin{figure}
  \centering
  \includegraphics[width=8cm]{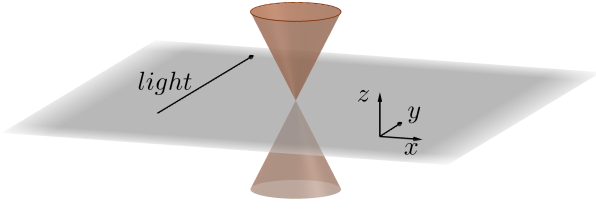}\\
  \caption{Schematic of a driven 2D system. The light travels in the y direction in the 2D plane. Before the driving is switched on, the 2D band structure has one Dirac cone. }\label{DiracInLight}
\end{figure}

The effective Hamiltonian in terms of Floquet theory reads $\mathcal{H}_{eff}=\mathcal{H}_0+[\mathcal{H}_1,\mathcal{H}_{-1}]/\omega+O(1/\omega^2)$\cite{3,6,11}, where $\mathcal{H}_n=\frac{1}{T}\int_0^T\mathcal{H}(t)e^{in\omega t}dt$ and $\mathcal{H}(t)=[k_x+A_x(t)]\sigma_x + i\partial_y\sigma_y$, because we chose $\mathbf{A}=(A_0\cos{(qy-\omega t)},0)$ in the graphene plane. Except $H_0$, the other terms in $H_{eff}$ vanish in this situation, which leaves the Dirac cones unchanged.

Instead, we choose to treat the external field through plane wave expansion, and express the field operators in terms of momentum representation\cite{23}
\begin{equation}\label{vector}
\begin{split}
  A_x&=A_0\cos(qy-\omega t)\\
     &=\frac{1}{2}A_0(e^{-i(qy-\omega t)}+e^{i(qy-\omega t)})\\
     &=N_q(ae^{-iqy}+a^\dagger e^{iqy})
\end{split}
\end{equation}
where $a$ ($a^\dagger$) is the annihilation (creation) operator of a photon with momentum $q$ and energy $w_q=q$, because we use the natural unit system of $e=c=\hbar=1$, and the normalization constant is defined as $N_q=\frac{A_0}{2\sqrt{\omega_q}}=\frac{A_0}{2\sqrt{q}}$. Thus, we can write the full Hamiltonian as
\begin{equation}\label{H}
\begin{split}
  H&=-i\int dr\Psi^\dagger(r)(\partial_i+iA_i)\sigma^i\Psi(r)+\omega a^\dagger a\\
   &=\sum_k[c_k^\dagger k_i\sigma^i c_k + N_q (a c_{k+q}^\dagger \sigma^x c_k +a^\dagger c_{k-q}^\dagger \sigma^x c_k)] +\omega a^\dagger a
\end{split}
\end{equation}
Here we use the Fourier transformation $\Psi(r)=\sum_k c_k e^{-ik \cdot r}$. The Hamiltonian is similar to the electron-phonon interaction in superconductors, if we replace the fermion operator with the two-component spinor.

\section{Effective interactions of fermions}

The first-order fermion-photon interaction can be cancelled out by Nakajima transformation, a unitary transformation $U=e^{-S}$ satisfying $S^\dagger=-S$ \cite{24}, which results in effective fermion-fermion interactions. $S$ can be written as
\begin{equation}\label{S1}
\begin{split}
  S&=\sum_k (ac_{k+q}^\dagger W c_k - a^\dagger c_{k-q}^\dagger W^\dagger c_k)\\
  W&=\beta_i \sigma^i ,  \quad i=0,x,y,z
\end{split}
\end{equation}
where $\beta_i$ are coefficients to be determined. The right $S$ can be determined by the condition
\begin{equation}\label{condition}
  H_I + [H_0,S] = 0
\end{equation}
where $H_0$ and $H_I$ are defined by
\begin{equation}\label{condition}
\begin{split}
  &H_0=\sum_kc_k^\dagger k_i\sigma^i c_k +\omega a^\dagger a\\
  &H_I=N_q (a c_{k+q}^\dagger \sigma^x c_k +a^\dagger c_{k-q}^\dagger \sigma^x c_k)
\end{split}.
\end{equation}
With $q=\omega$, one immediately get $\beta_x=\frac{N_q}{q}$. The other $\beta$ coefficients are complicated, but they will be neglected because they have no contribution. Actually, when considering the higher-order term $H_I^{eff}=\frac{1}{2}[H_I,S]$, one will immediately find that $\beta_y\sigma^y$ and $\beta_z\sigma^z$ terms will contribute nothing. As for the $\beta_0$ term, we can investigate it as follows.
With $\beta_x=\frac{N_q}{q}$, one can obtain the effective interaction
\begin{equation}\label{eff_int}
  H_I^{eff}=\frac{N_q^2}{q} \sum_{k,k'} c_{k+q}^\dagger \sigma^x c_k c_{k'-q}^\dagger \sigma^x c_{k'}
\end{equation}
Let $|M(k)\rangle=(e^{-i\phi(k)}, -1)^T/\sqrt{2}$ and $|P(k)\rangle=(e^{-i\phi(k)}, +1)^T/\sqrt{2}$ be the eigen vectors of the Hamiltonian $H_D$ for $-\epsilon$ and $+\epsilon$, respectively, where $\tan{\phi}=k_y/k_x$. Because the bands of $H_D(k)$ and $H_D(k+q)$ cross at $k_y=-q/2$ and the scattering $c_{k+q}^\dagger \sigma^x c_k$ mainly occurs at this $k$ point, it is reasonable to set the mean field $\Delta=\langle M(0,q/2)| \sum_k c_{k+q}^\dagger \sigma^x c_k |M(0,-q/2)\rangle=i$. Then, $\langle M| \sum_k c_{k+q}^\dagger c_k |M\rangle$ will be zero, and we can neglect the $\beta_0$ term. Therefore, the final effective Hamiltonian can be expressed as
\begin{equation}\label{H_eff}
  H_{eff}=\sum_k [c_k^\dagger \mathbf{k}\cdot\sigma c_k +(N_q^2\frac{\Delta}{q} c_{k+q}^\dagger \sigma^x c_k + h.c.)]
\end{equation}

\section{Band structure and topological property}

Noticing that the external field reduces the continuous (infinitesimal) translational symmetry to the discrete translational symmetry $\mathcal{T}: y \rightarrow y+2\pi/q$, the Brillouin zone will fold from infinity to $q$.  The band structure of (\ref{H_eff}) with $k_x=0$ is illustrated in Figure 2. The effective scattering in (\ref{H_eff}) will not make the Dirac point gapped, but will bend the band at $k=\pm\frac{q}{2}$. That makes the lowest two bands in the first Brllouin zone have a single-crossing feature, as illustrated in Figure 3. This is exactly an example of the Mobius twisted band discussed by several groups\cite{25,26}. The crossing of the bands is symmetry protected topological property by nonsymmorphic symmetry $G$ combined with inversion symmetry $P$\cite{26}, and the effective two-bands Hamiltonian is
\begin{equation}\label{H_eff2}
  \mathcal{H}(k_y)=\sin{(\frac{2\pi}{q}k_y)} \sigma^x +(1-\cos{(\frac{2\pi}{q}k_y)})\sigma^y
\end{equation}
The topology in this Hamiltonian can be understood in this way. For a one-dimensional Hamiltonian $\mathcal{H}(k)=h_x(k)\sigma^x+h_y(k)\sigma^y$, one can always choose a unitary matrix $U=e^{i\alpha(k) \sigma^y}e^{i\frac{\pi}{4}\sigma^x}$ to diagonalize the Hamiltonian with the rotational parameter $\tan{2\alpha(k)}=\frac{h_x(k)}{h_y(k)}$. In fact, both $U$ and $e^{i\frac{\pi}{2}\sigma^y}U$ diagonalize the same Hamiltonian. Thus, one can construct a fiber bundle $E$ with the base manifold of $S^1$ and the fiber of the rotating operator $e^{i\alpha(k) \sigma^y}e^{i\frac{\pi}{4}\sigma^x}$ and the structure group $\{I,e^{i\frac{\pi}{2}\sigma^y}\}$. The unitary matrixes $U(k)$ that diagonalize the Hamiltonian belong to one section of this fiber.

\begin{figure}
  \centering
  \includegraphics[clip,width=6cm]{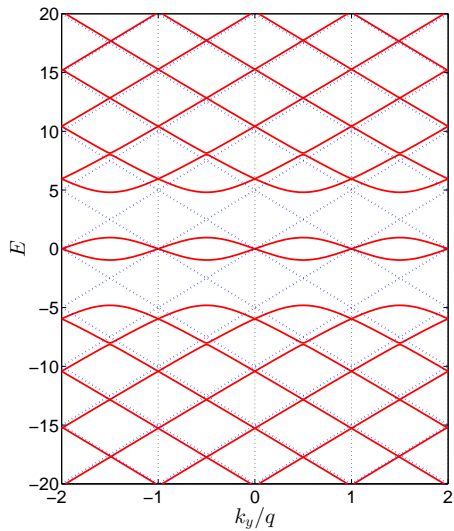}\\
  \caption{The band structure in the $k_y$ direction with $k_x=0$. The band structure describing $c_k$ fermions will describe $c_{k+q}$ fermions after a transform $k\rightarrow k+q$. The blue dot lines describe the initial Dirac fermions without driving, and the red lines are bands after the effective interactions are taken into account. }\label{EnergyInYdirection}
\end{figure}

To elucidate its topological property, we can divide $S^1$ into two parts of $[-\pi,0]$ and $[0,\pi]$, and construct $U(k)$ respectively, and thereby describe the topology by the homotopy group of the map from the overlapping zone $\{-\pi,0\}$ to the structure group $\{I,e^{i\frac{\pi}{2}\sigma^y}\}$. Consequently, the topology is characterized by $Z_2$. For the Hamiltonian (\ref{H_eff2}), one can easily show that it is topologically nontrivial. Intuitively speaking, with $k$ moves in $S^1$, $U(k)$ changes according to $\mathcal{H}(k)$, but when $k$ comes back to the starting point after a circle, $U(k)$ accumulates a factor $e^{i\frac{\pi}{2}\sigma^y}$, which will switch the two eigenstates of $\mathcal{H}(k)$. This means the energy levels of two eigenstates will interchange after a circle of $k$ movement, and guarantees the odd times of the band-crossings in a Brillouin zone, as illustrated in Figure 3.

\begin{figure}
  \centering
  \includegraphics[clip,width=6cm]{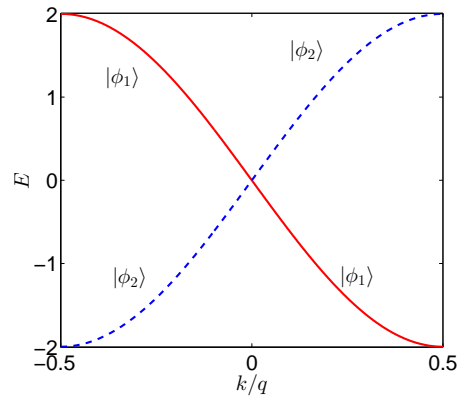}\\
  \caption{The band crossing for Hamiltonian (\ref{H_eff2}). The two eigenstates (and the energy levels) will interchange after a $2\pi$ period, which guarantees the odd times the number of the band crossing in the Brillouin zone. }\label{BandCross}
\end{figure}

\section{Energy splitting of Dirac cones}

\begin{figure}
  \centering
  \includegraphics[clip, width=8cm]{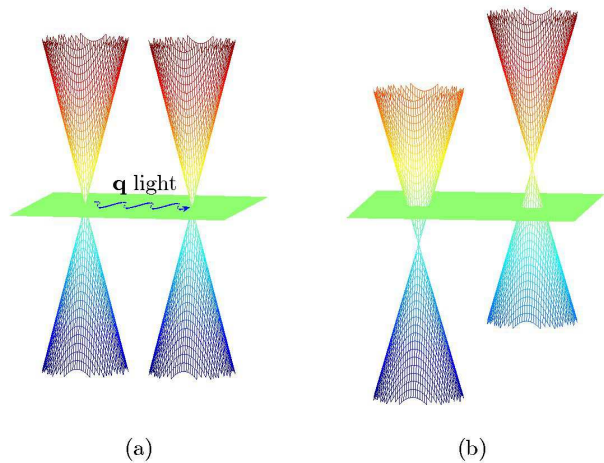}\\
  \caption{The lifting of the degenerate energy levels of the two Dirac cones, with circularly polarized light travelling along the $k_y$ direction. (a) The original degenerate two Dirac points described by the semi-Dirac Hamiltonian (\ref{semi}), separated in the $k_y$ direction. (b) The splitting of the energy levels of the two Dirac points due to the photon-mediated interactions. }\label{DiracSplit}
\end{figure}

It will become more interesting when we consider a system with two topologically different Dirac points, characterized by the chirality $W_D=sgn[det(V)]$ for the Hamiltonian $\mathcal{H}(k)=V_{ij}k_i\sigma^j$ for a Dirac cone. Let us consider a shift semi-Dirac-point Hamiltonian,
\begin{equation}\label{semi}
  H_{SD}=\sum_k c_k^\dagger [k_x\sigma^x + (k_y^2-b^2)\sigma^y] c_k
\end{equation}
This Hamiltonian describes two different Dirac points located at the two points: ($0,\pm b$). One can follow the same procedure  and find out that the $\beta_x$ values are determined by the following condition.
\begin{equation}\label{conditionofbeta}
\begin{split}
   -i[\frac{4k_x^2q^2}{\diamondsuit(\Box^2)+q^2}+\frac{q^2}{\diamondsuit}+\diamondsuit]\beta_z&=N_q\\
   (\frac{4k_x^2q}{\Box^2+q^2}+q)\beta_z+i\diamondsuit\beta_x&=0
\end{split}
\end{equation}
where the symbols are defined by $\Box=2k_yq+q^2$ and $\diamondsuit=(k_y+q)^2+k_y^2-2b^2$. Despite the complicated forms, we can still reduce it to a low-energy problem. Letting $k_x=0$ in (\ref{conditionofbeta}), we get the expression $\beta_x=\frac{qN_q}{\diamondsuit^2-q^2}$ and the effective Hamiltonian
\begin{equation}\label{H_eff3}
  H_{eff}=\sum_kc_k^\dagger [k_x\sigma^x + (k_y^2-b^2)\sigma^y] c_k +(N_q\Delta\beta_x c_{k+q}^\dagger \sigma^x c_k + h.c.)
\end{equation}
To study the low-energy physics and the the scattering from one Dirac point to the other, letting $k_y=-b$ and $q=2b$ in $\diamondsuit$, we shall get the scattering term $\frac{\Delta}{-q}\sigma^xc_{k+2b}^\dagger c_k+h.c.$. Combining with the free Dirac term, one can use a four-component $\Gamma$ matrix to describe the scattering
\begin{equation}\label{semi_eff}
  \mathcal{H}_{eff}^{(q=2b)}=k_x\sigma^x + 2bk_y\sigma^y\tau^z + \frac{N_q^2}{q}\sigma^x\tau^y
\end{equation}
where the Pauli matrixes $\tau$ describe the two Dirac points. This Hamiltonian can be easily bulk diagonalized by the unitary transformation $U=e^{i\frac{\pi}{4}\sigma^x\tau^x}$, and then we get
\begin{equation}\label{semi_eff_diag}
  \mathcal{H}_{eff}^{q=2b}=k_x\sigma^x + 2bk_y\sigma^y\tau^z + \frac{N_q^2}{q}\tau^z
\end{equation}
This means that the energy levels of the two Dirac points shift up and down by $\frac{A_0^2}{4q^2}$ respectively, as illustrated in Figure \ref{DiracSplit}, with the normalization constant $N_q=\frac{A_0}{2\sqrt{q}}$. The energy spacing is proportional to the square of the light strength $A_0$. Because $q=2b$ is decided by the irradiated material with two Dirac points, one can control the density of states  at the Fermi level by splitting the Dirac points through adjusting light strength, and thus control the transport property of the material. For the experimental lattice constant in a Dirac-cone system such as graphene ($a\approx1.4$\AA), the distance between two Dirac cones is around $2b\approx1.7$\AA{}$^{-1}$. This implies that the $X$-ray is needed, with a wave length of $\lambda\approx 3.7$\AA{}, because its wave vector is defined as $q=2\pi/\lambda$. Several recent experiments illustrated some honeycomb-like superstructures, with the spatial period being ten to hundred times the crystal constant, and thus they can create much closer Dirac cones in the first Brillouin zone\cite{27,28,29}. On the other hand, it is suggested that applying strain on graphene will destroy its $C_3$ symmetry and make its Dirac cones close to each other\cite{30}, and more candidates can be made available.

\section{Conclusion}

In summary, instead of usual Floquet method, we choose to treat the irradiated two-dimensional Dirac systems through capturing the photon-mediated electronic interactions. This method is especially important when we apply the light in the 2D plane of the Dirac systems. According to the perturbation theory in the quantum field theory, the mass shell condition combined with momentum conservation forbids the first-order procedure: absorption or emission of a photon. To capture the effective electronic interactions, we use Nakajima transformation to counteract the first-order electron-photon interaction, and thereby derive the effective electron-electron interactions mediated by photon. The effective interaction can result in interesting phenomena. For a Dirac system with one Dirac cone, it is found that the photon-mediated electronic interactions lead to the scattering at the Brillouin zone boundary and thus bend the bands there, and then produce the effective topologically nontrivial band structure protected by nonsymmorphic symmetry combined with inversion symmetry. For graphene, we find that the Dirac points at $K$ and $K^{'}$ will scatter from one to the other when the light wave vector $q$ matches the  distance between the Dirac points, and thus split them in energy level, i.e. lift the energy degeneracy of the Dirac points. The energy spacing is equivalent to $\frac{A_0^2}{2q^2}$, suggesting that one can control Dirac points through light strength to change the transport property of the Dirac system. This method should applicable to other systems, capturing their photon-mediated electronic interactions.

\begin{acknowledgments}
This work is supported by the Nature Science Foundation of China (Grant No. 11574366), by the Strategic Priority Research Program of the Chinese Academy of Sciences (Grant No.XDB07000000), and by the Department of Science and Technology of China (Grant No. 2016YFA0300701).
\end{acknowledgments}

\end{document}